\documentclass{PoS}

\title{The NEXT double beta decay experiment}

\ShortTitle{The NEXT double beta decay experiment}

\author{\speaker{Paola Ferrario}%
\thanks{On behalf of the NEXT Collaboration.}\\
\\
      Instituto de F\'isica Corpuscular (University of Valencia and CSIC) \\
      E-mail: \email{paola.ferrario@ific.uv.es}}

\abstract{NEXT (Neutrino Experiment with a Xenon TPC) aims to observe the neutrinoless double beta decay of \ensuremath{{}^{136}\rm Xe}  in a high-pressure gas xenon Time Projection Chamber using electroluminescence to amplify the signal from ionization. The two main advantages of this technology are a high energy resolution and the possibility of reconstructing electron tracks.

NEXT-100 is an electroluminescent, asymmetric TPC which will host 100 kg of the \ensuremath{{}^{136}\rm Xe}  isotope at 15 bar of pressure. On one side, a sparse array of photomultipliers records both the primary scintillation signal, which gives the starting time of the event, and electroluminescence, which gives a precise measurement of the total deposited energy. On the other side, a dense grid of silicon photomultipliers provides the reconstruction of the electron tracks. Being able to reconstruct the position of a track is doubly useful: on the one hand, it allows the correction of the energy of the event, which varies according to position, and on the other hand it provides an extra handle for background rejection, since a two-electron track shows higher energy density at both ends, while a single-electron track only at one end.

After a prototyping period (2009-2014) NEXT has completed the construction and started the operation of its first phase (NEW) in the Laboratorio Subterr\'aneo de Canfranc, in the Spanish Pyrenees, with the objectives of measuring the NEXT background model and the two-neutrino mode of the double beta decay.}

\FullConference{The European Physical Society Conference on High Energy Physics\\
		 5-12 July\\
		 Venice, Italy}

\begin{document}

\section{The NEXT experiment}

NEXT (Neutrino Experiment with a Xenon TPC) searches for neutrinoless double beta decay in a high pressure \ensuremath{{}^{136}\rm Xe} TPC at the Laboratorio Subterr\'aneo de Canfranc (LSC). 

The signal of a neutrinoless double beta decay is a peak in the kinetic energy spectrum of the outgoing electrons  at $Q_{\beta\beta}$. For this reason an experiment must be optimised simultaneously for energy resolution and the rejection of background events with energies near to $Q_{\beta\beta}$.  

Compared to liquid, gaseous xenon has a much lower fluctuation in the production of ionization charge, which results in more than ten times better intrinsic energy resolution for the ionization signal only \cite{intrinsic}. Also, charged particles leave extended tracks in gas, which allows one to veto alphas, electrons and muons coming from the outside. Therefore, the main background for this kind of detector comes from high energy gammas from environmental radioactivity entering the active volume. When gammas interact with xenon gas, they can produce photoelectric and Compton electrons, at energies very similar to Q$_{\beta\beta}$, which can mimic the signal \cite{Martin-Albo:2015rhw}. Electrons moving through xenon gas lose energy at an approximately fixed rate until they become non-relativistic. At the end of the trajectory the 1/$v^2$ rise of the energy loss (where $v$ is the speed of the particle) leads to an energy `blob', i.e., a high energy deposition in a compact region. This feature can be used to distinguish background single-electrons (one `blob' only at one extreme) from signal double-electrons (a single track with two `blobs' at the end-points). 

A charged particle propagating in the gas deposits its energy through both excitation and ionization of the gas molecules. The scintillation light coming from the relaxation of the molecules (UV photons at $\sim$ 172 nm) is registered by photomultipliers (PMTs) on the cathode side of the TPC and gives the starting time of the event. The ionization electrons are drifted by an electric field all the way through the drift region until they enter a small region of moderately higher field where they are accelerated and secondary scintillation (but not ionization) occurs. This process, called electroluminescence (EL), results in an amplification of the signal, which grows linearly with the electric field as long as its magnitude remains below the ionization threshold. PMTs detect the EL light, giving a precise measurement of the energy of the event. On the anode side, the distribution of the EL light on a grid of silicon photomultipliers (SiPMs) placed behind the EL area gives a 2D picture of the track at a given position along the axis, every microsecond. Given the starting time of the event, the absolute position along the TPC axis can also be reconstructed.

In a first phase NEXT has proven an excellent energy resolution in two prototypes, one at LBNL and the other one at IFIC. An energy resolution of 0.5\% FWHM extrapolated to the Q$_{\beta\beta}$ of \ensuremath{{}^{136}\rm Xe}, i.e., 2.458 MeV) has been measured \cite{dbdm} and the power of the two versus one `blob' signature for background rejection has been demonstrated \cite{Ferrario:2015kta}. 

\section{The NEW detector}

The NEXT-White\footnote{Named after Prof.~James White, our late mentor and friend.} (NEW) detector is a prototype of $\sim$50 cm of drift length, which contains about 5~kg of xenon mass in the active volume at 15 bar. Its purposes are to validate the technology choices for the final, 100 kg detector (NEXT-100), and to measure the background and the two-neutrino mode of the double beta decay.

\begin{figure}[htb]
\begin{center}
\includegraphics[width=0.5\textwidth]{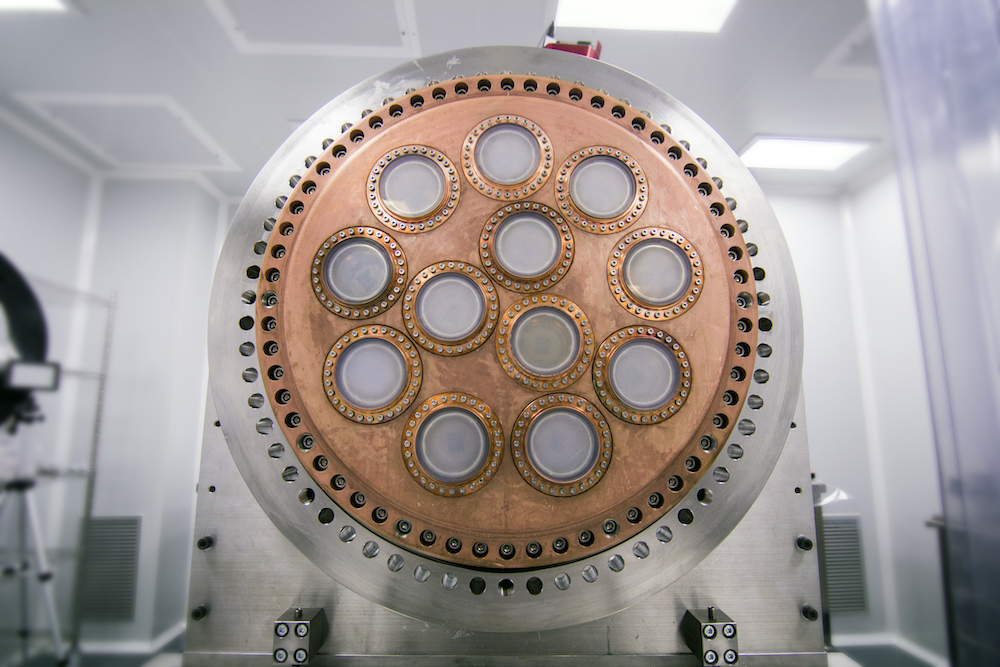}
\includegraphics[width=0.45\textwidth]{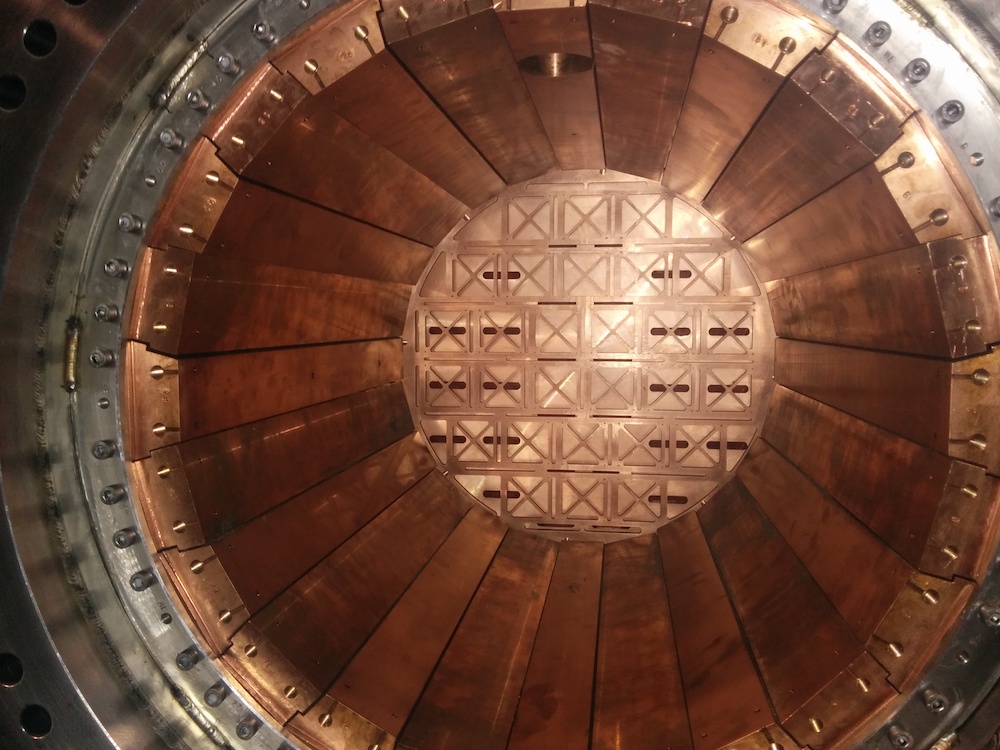}
\caption{Left: The NEW PMT plane. Right: the inner copper shield inside the NEW vessel.}
\label{fig:new}
\end{center}
\end{figure}

NEW consists of a cylindrical stainless-steel vessel, designed to withstand more than 20 bar of pressure, mounted on a seismic pedestal and surrounded by a lead castle (20~cm thick) that shields the detector against the high-energy gamma flux from the rocks of the laboratory. Inside the vessel a 6-cm thick radiopure copper shield protects the active volume from the radiation coming from the outside, including the vessel itself (see Fig.~\ref{fig:new}-\emph{right}). The readout planes are supported by 12-cm thick copper plates, and consist of 12 Hamamatsu R11410-10 photomultipliers (PMTs) and 1792 SensL C series silicon photomultipliers (SiPMs). The PMTs (see Fig.~\ref{fig:new}-\emph{left}) are 3-inch diameter and radiopure, with good quantum efficiency in the VUV and blue regions. The resulting photocathode coverage of the energy plane is about 30\%. 
The SiPMs of the tracking plane are mounted on 28 Kapton boards, each with $8\times8$ sensors separated by 1~cm with the nonsensitive areas covered with a reflective teflon layer to improve the light collection on the PMTs. The plane is positioned behind a 3-mm thick quartz plate coated with a conductive Indium Tin Oxide layer, which constitutes the anode. The quartz plate is also coated with TPB, since SiPMs are not sensitive to VUV light. The plate defines the ground end of the EL region, the other end being a stainless steel mesh
, kept at negative voltage. The field cage is made of a high density polyethylene tube, which holds copper rings designed to shape the electric field. Inside the rings, a 1-cm thick teflon cylinder, coated with TPB, serves as a reflector of light. The NEW field cage creates a homogeneous electric field of 300 $\textrm{V}~\textrm{cm}^{-1}$ in the active volume, and a field of 2-3 $\textrm{kV}~\textrm{cm}^{-1}~\textrm{bar}^{-1}$ in the EL region. 

\section{Calibration runs}

NEW has been running in calibration mode since October 2016 with depleted xenon, showing a high operational stability. 
Over the course of the runs the detector has been calibrated using a number of radioactive sources: \ensuremath{{}^{83}\rm Kr}  for geometric calibration and \ensuremath{{}^{22}\rm Na} and  \ensuremath{{}^{56}\rm Co} for medium to high energy calibration.

\begin{figure}[htb]
\begin{center}
\includegraphics[width=0.5\textwidth]{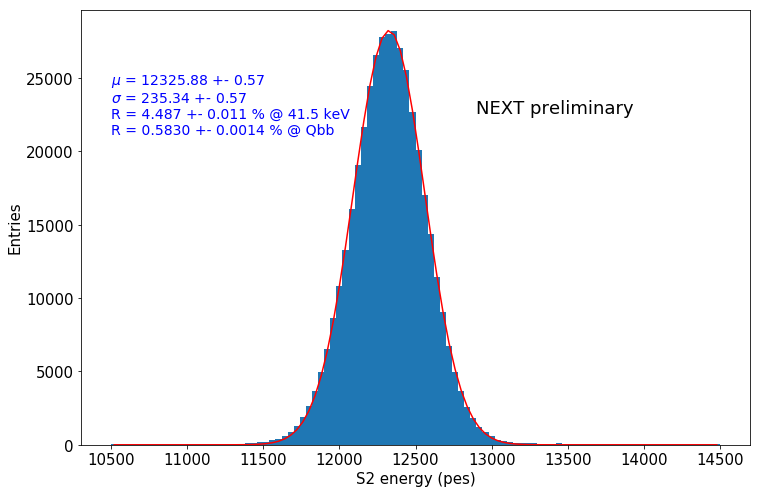}
\includegraphics[width=0.45\textwidth]{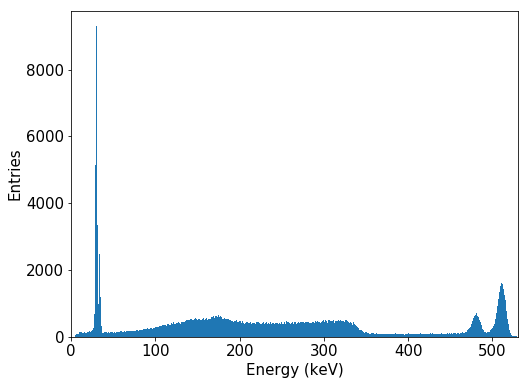}
\caption{Left: energy peak of a \ensuremath{{}^{83}\rm Kr} source, after corrections and applying a fiducial cut of radial coordinate less than 16 cm and drift coordinate within 5 and 45 cm. The extrapolation of the energy resolution to $Q_{\beta\beta}$ is done assuming statistical dependence only. Right: energy spectrum of a \ensuremath{{}^{22}\rm Na} source. The 511-keV photoelectric peak and the escape peak are clearly visible.}
\label{fig:res}
\end{center}
\end{figure}

\ensuremath{{}^{83}\rm Kr} is a metastable state of Kr, which decays emitting electrons that deposit their energy (about 41 keV in total) in a very small region of the gas (few mm in extent). The krypton is produced by the decay of \ensuremath{{}^{83}\rm Rb} zeolites inserted in the gas system, and diffuses rapidly, distributing uniformly throughout the whole chamber.  Krypton is very useful for characterizing the detector, in particular for measuring the drift velocity and mean lifetime of the electrons in the gas, both of which are affected by the level of impurities. In addition, by producing an energy deposition in a very small region, one can measure  the geometric dependence of the light detection precisely, allowing for an improvement of the energy resolution  to the levels required by the experiment. Several different data acquisition runs have been analyzed to perform a stable characterization of the detector, obtaining a drift velocity of 0.97 millimeters per microsecond and a mean electron lifetime of 1.5 milliseconds (increasing with time thanks to the continuous recirculation of the gas).  
Correcting for attachment and geometry, a resolution of  $4.5\%$ FWHM for the krypton peak has been found, as shown in Fig.~\ref{fig:res}-\emph{left}. 

Calibration runs  with a \ensuremath{{}^{22}\rm Na} source placed in one of the ports have also been taken, in which an external NaI scintillator was used to tag the backward 511-keV gamma for the trigger. The resulting energy spectrum has been corrected for lifetime and geometry with the same map extracted from Kr runs, and is shown in Fig.~\ref{fig:res}-\emph{right}. 

%
%

\begin{figure}[htb]
\begin{center}
\includegraphics[width=0.45\textwidth]{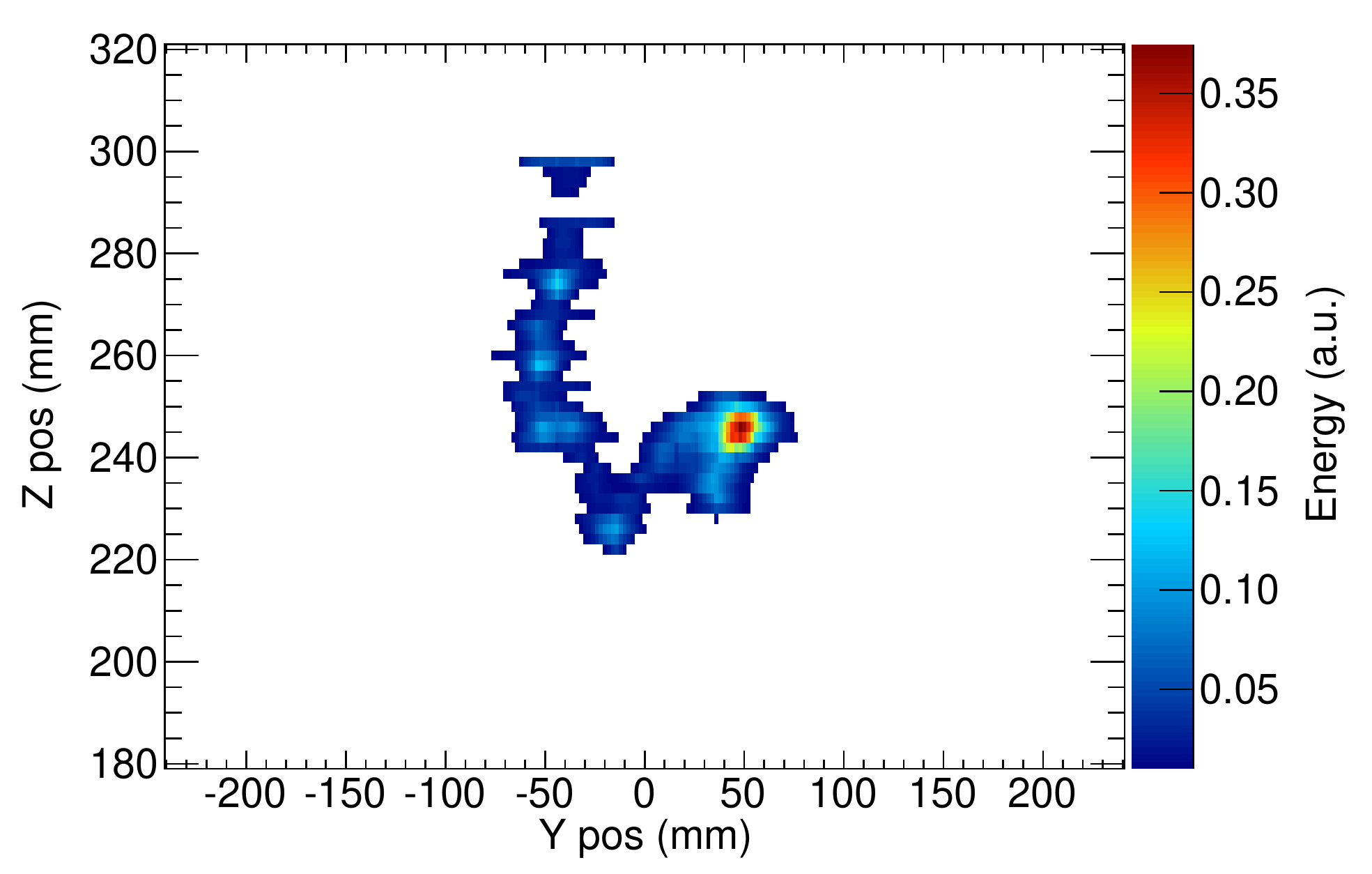}
\includegraphics[width=0.45\textwidth]{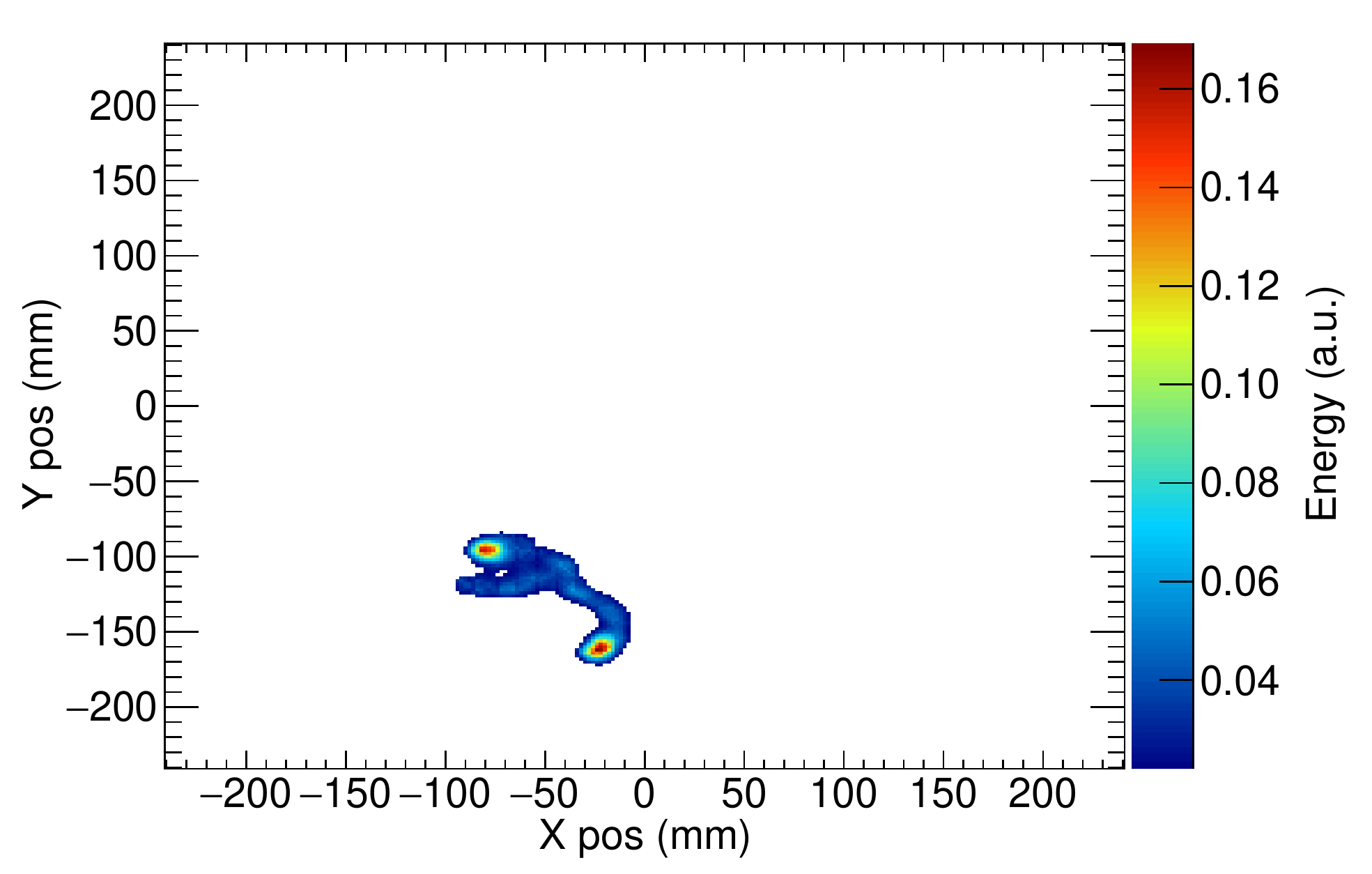}
\caption{Single electron (left) and electron-positron (right) tracks from a  \ensuremath{{}^{56}\rm Co} source, reconstructed with a ML-EM algorithm \cite{Simon:2017pck}.}
\label{fig:topo}
\end{center}
\end{figure}

Finally,  a \ensuremath{{}^{56}\rm Co}  source has been used to study the topological reconstruction and its rejection power. Since \ensuremath{{}^{56}\rm Co}  has several lines above the pair-production threshold, electron-positron pairs can be produced at several energies. Such tracks are topologically identical to a two electron signal in NEXT and, therefore, can be used to test the power of the the detector capability of distinguishing between single and double electron events. 
In Fig.~\ref{fig:topo} an example of single electron and $e^{+} e^{-}$  tracks are shown, where the one versus two energy `blobs' at the extremes are clearly distinguishable. 

\section{Conclusions}
In the NEXT experiment we search for neutrinoless double beta decay with gaseous xenon, a promising technology which combines an excellent energy resolution, signal-background discrimination based on tracking, and scalability. The first phase of NEXT, the NEW detector, has been running at LSC in a stable way for a year, demonstrating energy resolution and tracking capabilities with calibration sources. In 2018, a first low background run with depleted xenon will be carried out, followed by the search for the two-neutrino mode with enriched xenon.

\acknowledgments
The NEXT Collaboration acknowledges support from: the European Research Council (ERC) under the Advanced Grant 339787-NEXT; the Ministerio de Econom\'ia y Competitividad of Spain and FEDER under grants CONSOLIDER-Ingenio 2010 CSD2008-0037 (CUP), FIS2014-53371-C04 and the Severo Ochoa Program SEV-2014-0398; GVA under grant PROMETEO/2016/120. Fermilab is operated by Fermi Research Alliance, LLC under Contract No. DE-AC02-07CH11359 with the United States Department of Energy.


\begin{thebibliography}{99}
\bibitem{intrinsic} A.~Bolotnikov and B.~Ramsey, \emph{The spectroscopic properties of high- pressure xenon}, Nucl. Inst. Meth. A \textbf{396} (1997), 360-370.
\bibitem{Martin-Albo:2015rhw} The NEXT Collaboration (J.~Mart\'in-Albo {\it et al.}), \emph{Sensitivity of NEXT-100 to neutrinoless double beta decay}, JHEP \textbf{05} (2016) 159.
\bibitem{dbdm} The NEXT Collaboration (V.~\'Alvarez {\it et al.}), \emph{Near-intrinsic energy resolution for 30-662 keV gamma rays in a high pressure xenon electroluminescent TPC}, Nucl. Inst. Meth. A \textbf{708} (2013) 101-114.
\bibitem{Ferrario:2015kta} The NEXT Collaboration (P.~Ferrario {\it et al.}), \emph{First proof of topological signature in the high pressure xenon gas TPC with electroluminescence amplification for the NEXT experiment}, JHEP \textbf{01} (2016) 104.
\bibitem{Simon:2017pck}   The NEXT Collaboration (A.~Sim\'on {\it et al.}),  \emph{Application and performance of an ML-EM algorithm in NEXT},  JINST {\bf 12}, no. 08, P08009 (2017).
\end{thebibliography}
\end{document}